\documentstyle[amsfonts]{article}
\begin{document}
\title{Experimental realization of any discrete operator}
 \author{K. Svozil\\
  {\small Institut f\"ur Theoretische Physik}  \\
   {\small University of Technology Vienna }     \\
   {\small Wiedner Hauptstra\ss e 8-10/136}    \\
   {\small A-1040 Vienna, Austria   }            \\
   {\small e-mail: svozil@tph.tuwien.ac.at}\\
   {\small www: http://tph.tuwien.ac.at/$\widetilde{\;}$svozil}\\
 {\scriptsize
 this file is at
 http://tph.tuwien.ac.at/$\widetilde{\;}$svozil/publ/every.tex}
   }
\maketitle

\begin{abstract}
A straightforward argument shows that, by allowing counterfactual
elements of physical reality, any {\em arbitrary} discrete
finite-dimensional operator  corresponds to an observable.
\end{abstract}

As can be readily verified, any matrix $A$ can be decomposed into two
self-adjoint components
$A_1, A_2$ as follows.
\begin{eqnarray}
&&A=A_1+iA_2\label{e-decom1}\\
&&A_1={1\over 2}(A+A^\dagger) =:\Re A,\label{e-decom2}\\
&&A_2=-{i\over 2}(A-A^\dagger)=:\Im A.\label{e-decom3}
\end{eqnarray}
This is an extension of the decomposition of a complex number into its
real and
imaginary component; e.g., for any $a\in {\Bbb C}$, $a=a_1+i a_2$, with
$a_1=(1/2)(a+a^\ast )=:\Re a\in {\Bbb R}$ and
$a_2=-(i/2)(a-a^\ast )=:\Im a\in {\Bbb R}$ if $A=a$ is a
$(1\times 1)$-matrix.
Since the trace is additive,
this decomposition preserves the usual definition of quantum expectation
values even for the case of non pure states.

Thus, if $A_1$ and $A_2$ would consistently be measurable, then $A$
would consistently be measurable. As could be expected, in quantum
mechanics, this is guaranteed only if their commutator vanishes; i.e.,
$$[A_1,A_2]=A_1A_2-A_2A_1=0.$$
Note that, if $A=A^\dagger$ is self-adjoint, the decomposition is
trivial; i.e.,
$A_1=A$ and
$A_2=0$.

Nevertheless, quantum mechanics could be extended to counterfactual
elements of physical reality, as suggested by the
Einstein-Podolski-Rosen (EPR) argument \cite{epr}.
EPR state that
``if, without in any way disturbing a system, we can predict with
certainty (i.e., with probability equal to unity) the value of a
physical quantity, then there exists an element of physical reality
corresponding to this physical quantity.''
Thereby, EPR make no difference between an observable which
is actually measured and one which could only be obtained by
reasoning.
The term {\em counterfactual} can be defined, in Max Jammer's words
\cite[pp. 9--10]{jammer-92}, as follows.
``In general, an argument is called ``counterfactual'' if it involves
a thought experiment the actual performance of which on a given system
is made impossible because the conditions necessary for performing this
experiment cannot be satisfied.''
A shorter definition is due to Roger Penrose.
\cite[p. 204]{penrose:94},
``$\ldots$ things that might have happened, although they did not
happen.''
Surely enough, as has been shown by Bell
\cite{bell-66} (with an explicit reference to Gleason's theorem
\cite{Gleason}) and by Kochen and Specker
\cite{kochen1}, the naive assumption of non contextual counterfactuals
gives rise to inconsistencies and therefore excludes a broad class of
hidden parameter theories \cite{peres,mermin-93}.

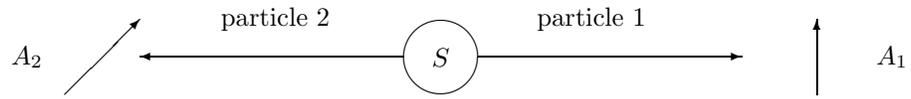
\begin{figure}[htd]
\begin{center}
\unitlength 1mm
\linethickness{0.4pt}
\begin{picture}(115.00,10.00)
\put(55.00,5.00){\circle{10.00}}
\put(55.00,5.00){\makebox(0,0)[cc]{$S$}}
\put(60.00,5.00){\vector(1,0){35.00}}
\put(50.00,5.00){\vector(-1,0){35.00}}
\put(105.00,0.00){\vector(0,1){10.00}}
\put(5.00,0.00){\vector(1,1){10.00}}
\put(115.00,5.00){\makebox(0,0)[cc]{$A_1$}}
\put(0.00,5.00){\makebox(0,0)[cc]{$A_2$}}
\put(75.00,10.00){\makebox(0,0)[cc]{particle 1}}
\put(33.00,10.00){\makebox(0,0)[cc]{particle 2}}
\end{picture}
\end{center}
\caption{\label{f-eprts}
EPR-type setup for measurement of arbitrary operator decomposed into its
self-adjoint real and imaginary part.
}
\end{figure}

With this {\it proviso}, one could nevertheless imagine
Gedankenexperiments which ``measure'' an arbitrary discrete observable,
irrespective of whether the corresponding operator is self-adjoint or
not.---Here, the quotes surrounding ``measure'' mean that
counterfactuals are involved. Assume, for instance, a spin one-half
system representable by twodimensional Hilbert space.
The matrix
$$A=
\left(
\begin{array}{cc}
0&0\\
1&0\\
\end{array}
\right)
$$
is not self-adjoint, but following equations
(\ref{e-decom1})--(\ref{e-decom3}) can be decomposed into two
self-adjoint operators
\begin{eqnarray}
&&A_1=
{1\over 2}
\left(
\begin{array}{cc}
0&1\\
1&0\\
\end{array}
\right)
={1\over 2}\sigma_1
,\label{e-decom2a}\\
&&A_2={1\over 2}
\left(
\begin{array}{cc}
0&i\\
-i&0\\
\end{array}
\right)
=-{1\over 2}\sigma_2
,\label{e-decom3a}\\
&&A=\sigma_1- i\sigma_2
.\label{e-decom4a}
\end{eqnarray}
Here, $\sigma_i, i=1,2,3$ denote the Pauli spin matrices.
In an EPR-type measurement setup drawn in Figure \ref{f-eprts}, two
entangled spin one-half particles in the singlet state could be
spatially separated and interrogated.
 Measurement on the first particle
could yield the element of physical reality
associated with $\sigma_1$;
that is the spin state along the $x-axis$.
Measurement on the second particle
could yield the element of physical reality
associated with $\sigma_2$;
that is the spin state along the $y-axis$.
These outcomes could then be combined
according to Equation (\ref{e-decom1}) to yield the element of physical
reality associated with $A$.

It is not difficult to imagine that, since we can always decompose an
arbitrary observable into just two self-adjoint observables, a
generalized EPR-type experiment combined with the
Reck--Zeilinger--Bernstein--Bertani setup
\cite{rzbb}, every discrete finite-dimensional operator which is not
necessarily
self-adjoint, is ``measurable'' (in the counterfactual sense).


\begin{thebibliography}{10}

\bibitem{bell-66}
John~S. Bell.
\newblock On the problem of hidden variables in quantum mechanics.
\newblock {\em Reviews of Modern Physics}, 38:447--452, 1966.
\newblock Reprinted in \cite[pp. 1-13]{bell-87}.

\bibitem{bell-87}
John~S. Bell.
\newblock {\em Speakable and Unspeakable in Quantum Mechanics}.
\newblock Cambridge University Press, Cambridge, 1987.

\bibitem{epr}
Albert Einstein, Boris Podolsky, and Nathan Rosen.
\newblock Can quantum-mechanical description of physical reality be considered
  complete?
\newblock {\em Physical Review}, 47:777--780, 1935.
\newblock Reprinted in \cite[pp. 138-141]{wheeler-Zurek:83}.

\bibitem{Gleason}
Andrew~M. Gleason.
\newblock Measures on a closed subspaces of a {H}ilbert space.
\newblock {\em Journal of Mathematics and Mechanics}, 6:885--893, 1957.

\bibitem{jammer-92}
Max Jammer.
\newblock John {S}teward {B}ell and the debate on the significance of his
  contributions to the foundations of quantum mechanics.
\newblock In A.~van~der Merwe, F.~Selleri, and G.~Tarozzi, editors, {\em Bell s
  Theorem and the Foundations of Modern Physics}, pages 1--23. World
  Scientific, Singapore, 1992.

\bibitem{kochen1}
Simon Kochen and Ernst~P. Specker.
\newblock The problem of hidden variables in quantum mechanics.
\newblock {\em Journal of Mathematics and Mechanics}, 17(1):59--87, 1967.
\newblock Reprinted in \cite[pp. 235--263]{specker-ges}.

\bibitem{mermin-93}
N.~D. Mermin.
\newblock Hidden variables and the two theorems of {J}ohn {B}ell.
\newblock {\em Reviews of Modern Physics}, 65:803--815, 1993.

\bibitem{murnaghan}
F.~D. Murnaghan.
\newblock {\em The Unitary and Rotation Groups}.
\newblock Spartan Books, Washington, 1962.

\bibitem{penrose:94}
Roger Penrose.
\newblock {\em Shadows of the Minds, A Search for the Missing Science of
  Consciouness}.
\newblock Oxford University Press, Oxford, 1994.

\bibitem{peres}
Asher Peres.
\newblock {\em Quantum Theory: Concepts and Methods}.
\newblock Kluwer Academic Publishers, Dordrecht, 1993.

\bibitem{rzbb}
M.~Reck, Anton Zeilinger, H.~J. Bernstein, and P.~Bertani.
\newblock Experimental realization of any discrete unitary operator.
\newblock {\em Physical Review Letters}, 73:58--61, 1994.
\newblock See also \cite{murnaghan}.

\bibitem{specker-ges}
Ernst Specker.
\newblock {\em Selecta}.
\newblock Birkh{\"{a}}user Verlag, Basel, 1990.

\bibitem{wheeler-Zurek:83}
John~Archibald Wheeler and Wojciech~Hubert Zurek.
\newblock {\em Quantum Theory and Measurement}.
\newblock Princeton University Press, Princeton, 1983.

\end{thebibliography}

\clearpage

\end{document}